\documentclass[letter,longauth, traditabstract]{aa} 
\usepackage{graphicx}
\usepackage{txfonts}
\usepackage{natbib}

\newcommand{\CSO}{\textrm{\CSO}}

\newcommand{\emm}[1]{\ensuremath{#1}}   
\newcommand{\emr}[1]{\emm{\mathrm{#1}}} 
\newcommand{\unit}[1]{\emm{\, \emr{#1}}}
\newcommand{\Msol}  {\unit{M_\odot}}

\renewcommand{\deg}{\emm{^\circ}}

\begin{document}
\title{The $^{35}$Cl/$^{37}$Cl isotopic ratio in dense molecular clouds:\\
HIFI observations of hydrogen chloride towards W3\,A\thanks{Herschel is an ESA space observatory with science instruments provided by European-led Principal
Investigator consortia and with important particiation from NASA.}}

\author{J. Cernicharo\inst{1}, J.R. Goicoechea\inst{1}, F. Daniel \inst{1}%
\and M. Ag\'undez\inst{1,2}
\and E. Caux \inst{3}
\and T. de Graauw \inst{4}
\and\\ A. De Jonge \inst{5}
\and D. Kester \inst{5}
\and H.G. Leduc \inst{6}
\and E. Steinmetz \inst{7}
\and J. Stutzki \inst{8}
\and J.S. Ward \inst{9}}

\offprints{\\\email{jcernicharo@cab.inta-csic.es}}

\institute{Centro de Astrobiolog\'{\i}a. CSIC-INTA.
Carretera de Ajalvir, Km 4, Torrej\'on de Ardoz. 28850, Madrid, Spain
\email{jcernicharo@cab.inta-csic.es}
\and
LUTH, Observatoire de Paris-Meudon, 5 Place Jules Janssen, 92190 Meudon, France.
\and
Centre d'Etude Spatiale des Rayonnements, Universit\'e  de Toulouse [UPS], 
31062 Toulouse Cedex 9, France
\and
Atacama Large Millimeter/Submillimeter Array, ALMA Office, Santiago, Chile
\and
SRON Netherlands Institute for Space Research, Landleven 12, 9747 AD Groningen
\and
Jet Propulsion Laboratory, 4800 Oak Grove Drive, MC 302-231, Pasadena, CA 91109  U.S.A.
\and
MPI f\"ur Sonnensystemforschung, D 37191 Katlenburg-Lindau, Germany
\and
KOSMA, I. Physik. Institut, Universit\"at zu K\"oln, Germany
\and
 with Raytheon Co., Fort Wayne, Indiana, U.S.A., since March of 2009}

\date{Received 30 March 2010 ; Accepted 12 May 2010}

\abstract{We report on the detection with the HIFI instrument on
board the Herschel satellite of the two hydrogen chloride
isotopologues, H$^{35}$Cl and H$^{37}$Cl, towards the massive 
star-forming region W3\,A. The $J$=1--0 line of both species
was observed with receiver \textit{1b} of the HIFI
instrument at $\sim$625.9 and $\sim$624.9\,GHz. The different
hyperfine components were resolved. The
observations were modeled with a non-local, non-LTE radiative
transfer model that includes hyperfine line overlap and radiative
pumping by dust. Both effects are found to play an important role in the
emerging intensity from the different hyperfine components. The
inferred H$^{35}$Cl column density (a few times
$\sim$10$^{14}$\,cm$^{-2}$), and fractional abundance
relative to H nuclei ($\sim$7.5$\times$10$^{-10}$), supports an
upper limit to the gas phase chlorine depletion of $\approx$200. Our
best-fit model estimate of the H$^{35}$Cl/H$^{37}$Cl abundance ratio
is $\approx$2.1$\pm$0.5, slightly lower, but still compatible with
the solar isotopic abundance ratio ($\approx$3.1). Since both species
were observed simultaneously, this is the first accurate estimation
of the [$^{35}$Cl]/[$^{37}$Cl] isotopic ratio in molecular
clouds. Our models indicate that even for large line
opacities and possible hyperfine intensity anomalies, the
H$^{35}$Cl and H$^{37}$Cl $J$=1-0 integrated line-intensity ratio
provides  a  good estimate of the $^{35}$Cl/$^{37}$Cl  isotopic abundance ratio.}


\titlerunning{The $^{35}$Cl/$^{37}$Cl ratio in molecular clouds}
\maketitle

\section{Introduction}
\label{sec:introduction}

Chlorine has two stable isotopes ($^{35}$Cl and $^{37}$Cl) and an
ionization potential of 12.97 eV (\textit{i.e.,} slightly below that of
hydrogen). Hence, it can be ionized by UV photons (912-956
\AA) in diffuse clouds and in the edges of photon-dissociation
regions (PDRs). Once ionized, Cl$^+$ reacts
 with molecular hydrogen exothermically to form HCl$^+$,
a process that initiates the chemical reactions of chlorine. In cloud
interiors, HCl$^+$ can be formed by reactions starting 
with neutral Cl and H$_3^+$. The chemistry of chlorine in interstellar
clouds has been the subject of various works (see Neufeld \& Wolfire 2009). 
These studies predict that hydrogen
chloride (HCl) is the most abundant Cl-bearing molecule in
dense clouds.

The HCl hyperfine lines (see Sect.~3.2) can be resolved in
interstellar sources only with heterodyne receivers equipped with
high spectral resolution spectrometers. The HCl $J$=1-0 line at
$\sim$625.9\,GHz was first detected with the \textit{Kuiper
Airbone Observatory} towards Orion \citep{Blake1985} and followed
by detections in Sgr~B2 \citep{Zmuidzinas1995}, several positions
in OMC-1 \citep{Schilke1995}, and in Mon~R2   using the
\textit{Caltech Submillimeter Observatory} with good atmospheric
transparency (Salez et al. 1996; SFL96 hereafter). The inferred HCl
column densities are in the range 10$^{13}$-10$^{14}$\,cm$^{-2}$.
SFL96 also presented the first detection of  H$^{37}$Cl  towards
Orion. Since the chemical reactions involving HCl are relatively
well understood (see Sect.~\ref{chemistry}), the
 H$^{35}$Cl/H$^{37}$Cl abundance ratio should provide a good
measure of the $^{35}$Cl/$^{37}$Cl isotopic ratio. Both $^{35}$Cl and
$^{37}$Cl nuclei are believed to  form in the last burning
stages of massive stars ($>$10\,\Msol) and by means of ``explosive
nucleosynthesis" during supernovae detonation
(\textit{e.g.,} Woosley \& Weaver 1995).
Therefore, observations of H$^{35}$Cl and H$^{37}$Cl, and accurate
measurements of the $^{35}$Cl/$^{37}$Cl ratio in different
environments, can provide some insight into the chemical
evolution of both isotopes, thus into Galactic chemical
evolution. The species HCl was also detected recently
towards the carbon-rich evolved star IRC+10216 by \citet{Cernicharo2010}.

Using HIFI, the \textit{Heterodyne Instrument for the
Far-Infrared} \citep{deGraauw2010}, on board the \textit{Herschel
Space Observatory} \citep{Pilbratt2010}, we present in this
\textit{Letter} the detection of the $J=1-0$ rotational
transition of H$^{35}$Cl and H$^{37}$Cl towards the massive 
star-forming region W3. The broad frequency coverage of
\emph{HIFI} allows us to observe both isotopologues of HCl with the
same relative calibration in a wide variety of
astronomical environments. Here we present the
first accurate determination of the $^{35}$Cl/$^{37}$Cl isotopic
ratio by a detailed model of the excitation of the hyperfine
levels developed using a non-local radiative transfer code. The effect of
line overlaps between the hyperfine components, and radiative
pumping by dust photons are discussed and modeled in
detail.

\section{Observations and data reduction}
\label{sec:obs}

All  spectra presented here were taken
during the performance verification (PV) phase of HIFI (de Graauw et al. 2010).
Both H$^{35}$Cl and H$^{37}$Cl $J$=1-0 lines were observed in the
\textit{Band~1b} receiver  using the Wide Band Spectrometer (WBS), which
provides $\sim$4\,GHz
of bandwidth and $\sim$1.1\,MHz of channel spectral resolution
(or a velocity resolution of $\sim$0.5\,km\,s$^{-1}$ at $\sim$626\,GHz).
The telescope, which has a half-power beam-width (HPBW) of   $\sim$35$''$
at $\sim$626\,GHz, was centered on
$\alpha_{2000} = 02^h25^m43.51^s$, $\delta_{2000} = 62\deg 06'13''$
(a position close to  W3\,A IRS\,2 and 2a).
A complete spectral scan of the \textit{Band~1b} was taken at this position
during PV phase and several lines from different molecules have been identified so far.
The horizontal (H) and vertical (V) polarization receivers were averaged after
rescaling the V one using the HCO$^+$ J=7-6 line intensity at 624.21\,GHz, i.e., 
close to that of HCl isotopologues.
The data were first processed with HIPE software (Ott et al. 2010), and then exported to CLASS
where standard data reduction routines were carried out.
We checked that the target lines are not contaminated by lines
from the other side band using the different frequency settings. 
The rms noise at $\sim$626\,GHz
is $\sim$30\,mK (antenna temperature) per 0.5\,km\,s$^{-1}$ resolution channel.
Hence, the  H$^{35}$Cl and H$^{37}$Cl $J$=1-0 lines are detected at  10$\sigma$ and
6$\sigma$ levels, respectively and with the same relative calibration. 
At this level of sensitivity, only the most abundant
species are detected. In particular, we  detected 20 lines 
in the entire band $1b$ above 3$\sigma$. The brightest is the CO $J$=5-4 line, followed by the 
ground-state line of ortho--H$_2$O and a few lines from formaldehyde, methanol, HCO$^+$, and HCN. 
Hence, we are confident that the observed HCl line profiles are not blended with  other
spectral features. By examining our own and public spectral catalogs 
\citep{Muller2001,Muller2005, Pickett1998}, we also checked that lines  
in the signal band from other molecules do not blend with the hyperfine components 
of both  HCl isotopologues.

\begin{figure}
\centering
\rotatebox{0}{
\includegraphics[width=0.49\textwidth]{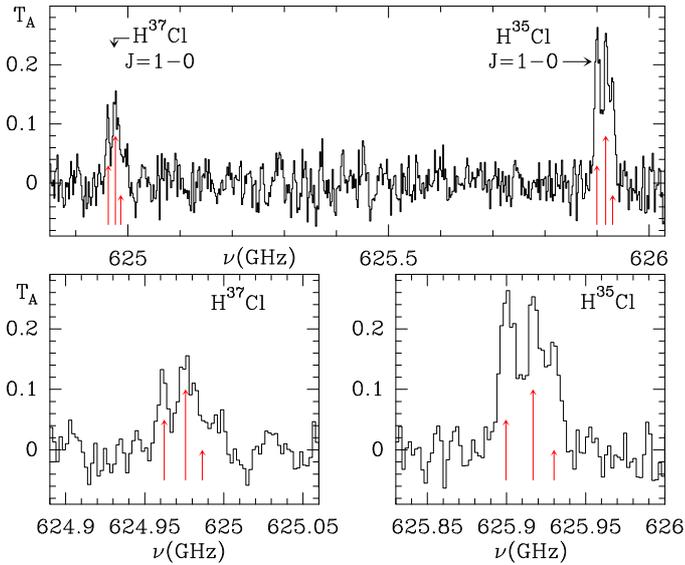}}
\caption{Detection of H$^{35}$Cl and H$^{37}$Cl $J$=1-0 lines towards W3\,A H{\sc ii} region.
Arrows shows the relative line strength of each HFS  component.
The length of the arrows are proportional to the expected intensities in the LTE
optically thin limit. Both lines were observed simultaneously in a line survey of band 
1b of HIFI, hence, have the same calibration accuracy. 
Spectral resolution was smoothed to 2 MHz ($\simeq$1 km\,s$^{-1}$).}
\label{fig:hcl_hifi}
\end{figure}

The total integration time 
was 2 minutes.
Figures~\ref{fig:hcl_hifi} 
and \ref{fig:w3a_hifi_lines} show the resulting line profiles
(data smoothed to a spectral resolution of $\simeq$1\,km\,s$^{-1}$).
To compare with 
our  models, the following expression for the main beam efficiency
was adopted,  $\eta_{mb} = 0.72\,exp(-(\nu/6)^2)\times0.96$, where
$\nu$ is the frequency in THz,
0.72 is $\eta_{mb}$ for the telescope in the limit of 0 frequency,
and the factor 0.96 is the assumed forward efficiency of the telescope
(M. Olberg priv. comm.).

\section{Results}
\label{sec:results}
In terms of spectroscopy, the $I$=3/2 nuclear spin of $^{35}$Cl and of $^{37}$Cl splits the
pure rotational transitions of H$^{35}$Cl and H$^{37}$Cl
into several hyperfine structure (HFS) components
(see \textit{e.g.,} Cazzoli \& Puzzarini 2004 and
references therein).
These hyperfine components are indicated as vertical arrows
in Fig.~\ref{fig:hcl_hifi}. In the optically thin limit, these components  follow
a 2:3:1 intensity ratio (from the lowest to the highest frequency hyperfine component). 
It is clear from Fig.~\ref{fig:hcl_hifi} that the observed ratios are close to 1:1:1 for the
three hyperfine components of H$^{35}$Cl. These ratios 
indicate that the H$^{35}$Cl hyperfine components are affected substantialy by opacity.
Even for H$^{37}$Cl, the observed
hyperfine line intensities (1:1:2) do not follow the expected ratios in an optically thin case.

\begin{figure}
\centering
\rotatebox{0}{
\includegraphics[width=0.48\textwidth]{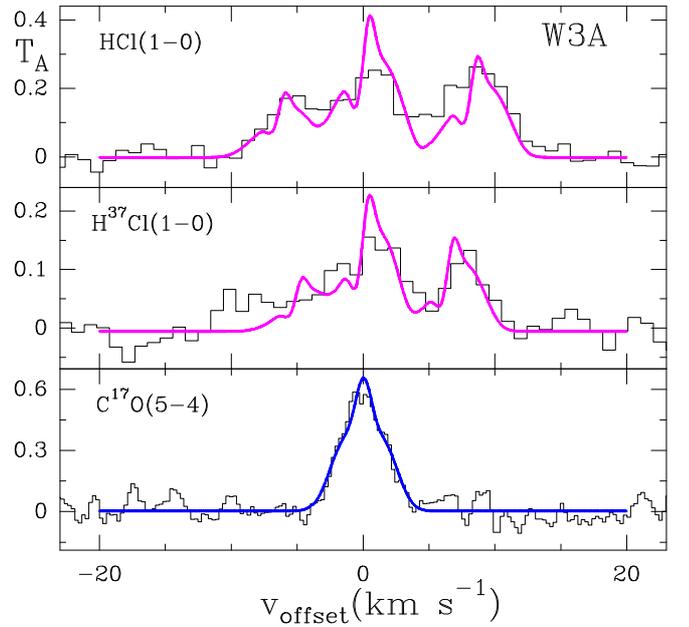}}
\caption{H$^{35}$Cl, H$^{37}$Cl $J$=1-0,  and C$^{17}$O~$J$=5-4 lines observed
with HIFI towards W3\,A.
Continuous curves show closest fitting radiative-transfer model line
profiles  (see text). 
The spectral resolution is $\simeq$1\,km\,s$^{-1}$} \label{fig:w3a_hifi_lines}
\end{figure}


The intensity peak ratio of the strongest
H$^{35}$Cl and H$^{37}$Cl HFS components is $\simeq$1.5 and
the integrated intensity ratio is $\simeq$2$\pm$0.2.
These values are lower than the solar $^{35}$Cl/$^{37}$Cl abundance
ratio, $\sim$3.1 \citep{Anders1989}, which suggests that optical
depth effects could influence the observed HFS line intensity ratios (or
that the H$^{35}$Cl/H$^{37}$Cl abundance ratio is lower than the
solar value). For completeness, Figure~\ref{fig:w3a_hifi_lines}
compares the detected HCl lines with the C$^{17}$O $J$=5-4 at 561.712\,GHz
also observed with HIFI.
The detection of  C$^{17}$O, HCO$^+$ and HCN mid-$J$ lines confirms the presence of warm and dense
molecular gas towards the observed position.

\subsection{HCl excitation and $^{35}Cl/^{37}Cl$ abundance ratio}
\label{column_densities}
The star-forming region W3 is located in the Perseus arm
at a distance of 2.3\,kpc
and contains several young massive stars that ionize a natal molecular cloud
creating H{\sc ii} regions.
In particular, the near-IR sources IRS\,2 and IRS\,2a (OB stars)
are believed to be the ionizing sources of the W3\,A  H{\sc ii} region
(Tieftrunk et al. 1995 and references therein). These sources are also the origin of molecular outflows
and are X--rays emitters \citep{Hofner2002}.

\begin{figure}
\centering
\rotatebox{-90}{
\includegraphics[width=0.22\textwidth]{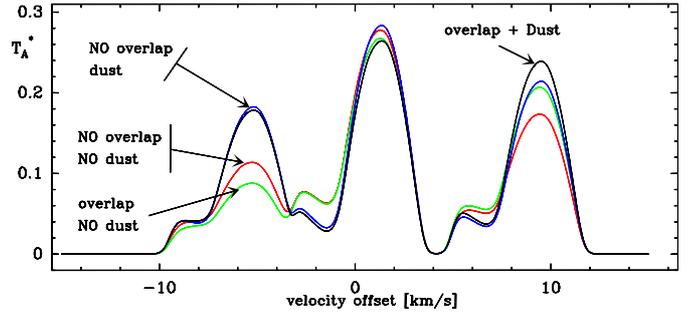}}
\caption{Results from non-local, non-LTE radiative
transfer calculations for the H$^{35}$Cl $J$=1-0 HFS components.
The different curves show the effects of including line overlap
and radiative pumping from dust photons in the emerging line
intensities and relative HFS line ratios.}
\label{fig:transfer-effect}
\end{figure}

To estimate the
H$^{35}$Cl/H$^{37}$Cl abundance ratio and  analyze all possible effects
affecting the emerging line profiles, we  modeled the observed
HCl and C$^{17}$O lines with our non-local and
non-LTE radiative transfer  codes \citep{Gonzalez1993,
Gonzalez1997, Goicoechea2006, Daniel2008}.
To take into account the blending of the HCl $J$=1--0 hyperfine
components (\textit{i.e.,} to determine the opacity at each
frequency when several lines overlap), we  used the
modelling approach presented in  Daniel \& Cernicharo
(2008) to interpret the HFS line emission from N$_2$H$^+$, HCN,
and HNC. Rate coefficients for the collisional excitation of HCl
by He are taken from Neufeld \& Green~(1994), who also 
estimated  the contribution to each specific hyperfine level.
Like other light hydrides that HIFI will observe, HCl has a large
rotational constant (10.4\,cm$^{-1}$),  thus HCl rotational
transitions have high spontaneous radiative rates. 
The HCl critical densities are very high,
n$_{cr}$(J=1-0)$\simeq$7$\times$10$^7$\,cm$^{-3}$, and only when
n(H$_2$)$\gtrsim$n$_{cr}$ does collisional excitation  dominate.
This high value suggests that HCl line emission arises in dense
molecular gas. Radiative pumping by dust photons
may also be very important in
determining the HCl level populations (see also SFL96) because of 
the increase in grain emissivity and dust opacity in the far-IR and
submillimeter domains \citep{Cernicharo2006a, Cernicharo2006b}.
Model predictions shown  in Fig.~\ref{fig:transfer-effect} demonstrate that the
inclusion of line overlap, dust pumping, and both effects
together, modifies the relative intensity of each HFS component.
We note in particular how the $J=1-0$\,\,\,$F$=1/2--1/2 line (the HFS
component with the weakest line strength and opacity) is enhanced
with respect to the other  components when radiative pumping is
included.

To reproduce the observed H$^{35}$Cl and  H$^{37}$Cl  line profiles and relative
intensities, we 
assumed uniform physical conditions
(n$_H$$\simeq$10$^6$\,cm$^{-3}$ and T$_k$$\simeq$100\,K;
taken from Helmich et al. 1994; Tieftrunk et al. 1995) and that
 the  H$^{35}$Cl and  H$^{37}$Cl  abundances
are free parameters. The C$^{17}$O line was also analyzed with the same parameters to
verify the model consistency. In particular, the C$^{17}$O $J$=5-4 line was found to be optically thin
($\tau$$\simeq$0.1), which allowed us to constrain the  line-of-sight column density of material
and also the line velocity dispersion ($\sigma$$\simeq$1.3\,km\,s$^{-1}$).
Gas and dust were assumed to coexist and be thermally coupled (T$_k$=T$_d$).

To reproduce the observed HCl line peak positions and their
relative  strengths, best fit solutions  were obtained for an expanding shell
of gas. The adopted velocity gradient (from 2.5\,km\,s$^{-1}$ at
the center  to 0.5\,km\,s$^{-1}$ at the edge) is consistent with
the CO molecular outflows seen in the region (\textit{e.g.,}
Hasegawa et al. 1994). Optimal results were obtained for a H$^{35}$Cl
column density of a few times 10$^{14}$\,cm$^{-2}$ (or an
abundance of $\sim$7.5$\times$10$^{-10}$ relative to total H).
The inclusion of radiative pumping from dust (and line overlap to a lesser extent)
allowed one to more accurately reproduce the observed HFS relative line
intensities. 
Assuming that  H$^{35}$Cl and C$^{17}$O arise from the same regions, we
inferred a column density ratio of $N$(H$^{35}$Cl) $\simeq$ $N$(C$^{17}$O)/20, by
using the CO abundance determined for the region ($\sim$4$\times$10$^{-5}$ per H nucleus; 
\citet{Tielens1991}) and assume a standard $^{16}$O/$^{17}$O isotopic ratio of 2600.
Optimal fits were obtained by using an automatic $\chi^2$
procedure and are shown in Fig.~\ref{fig:w3a_hifi_lines}. The
H$^{35}$Cl/H$^{37}$Cl abundance ratio found in the models is
$\sim$2.1 (with a confidence interval within 1.6--3.1). Since
H$^{35}$Cl lines are moderately optically thick ($\tau$$\sim$12, 8, 4 for each HFS component), 
one does not expect the
observed H$^{35}$Cl/H$^{37}$Cl line intensity ratio to provide a direct measure of the
H$^{35}$Cl/H$^{37}$Cl abundance ratio. However, HCl critical densities are much higher
than the gas density in most ISM clouds, and therefore  the excitation temperature 
of the different HFS components remain proportional to the HCl column density,
even for optically thick lines. 
To conclude whether or not the line integrated intensity ratio is a good measure
of the isotopic abundance ratio, Fig.~4 shows the modeled 
integrated intensity ratio as a function of the H$^{35}$Cl/H$^{37}$Cl abundance ratio. Although
the hyperfine components are optically thick in most models, the
integrated line intensity ratio is proportional to the isotopic ratio for
HCl abundances below 10$^{-9}$. In Fig.~4, we also show the results for a static
cloud. In this case, the opacities are larger but the integrated intensity ratio still 
provides a reliable
measurement of the isotopic abundance ratio for HCl abundances below 10$^{-10}$.

\begin{figure}
\centering
\includegraphics[width=0.35\textwidth,angle=-90]{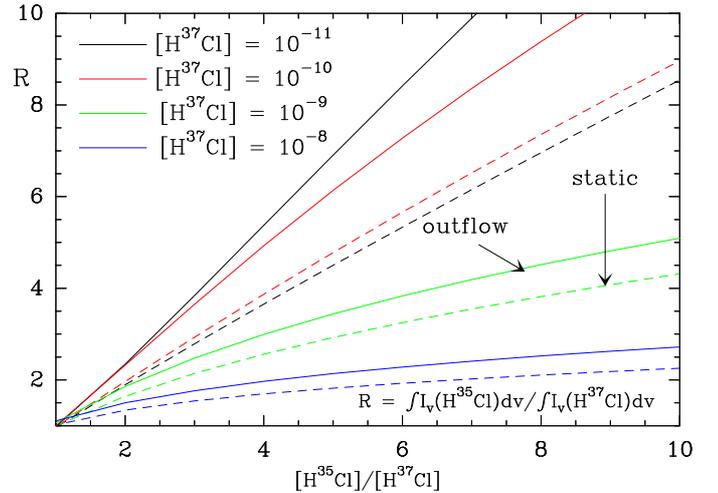}
\caption{
Modeled HCl and H$^{37}$Cl $J$=1-0 integrated line intensity ratio ($R$)
as a function of the assumed isotopic abundance ratio (see text). 
HCl abundances, indicated at the top left, are relative to H nuclei. 
Continuous lines correspond to models with a velocity gradient. 
Dashed lines correspond to a static model.}
\label{fig:ratio-model}
\end{figure}

\begin{figure}
\centering
\rotatebox{-90}{
\includegraphics[width=0.35\textwidth,angle=0]{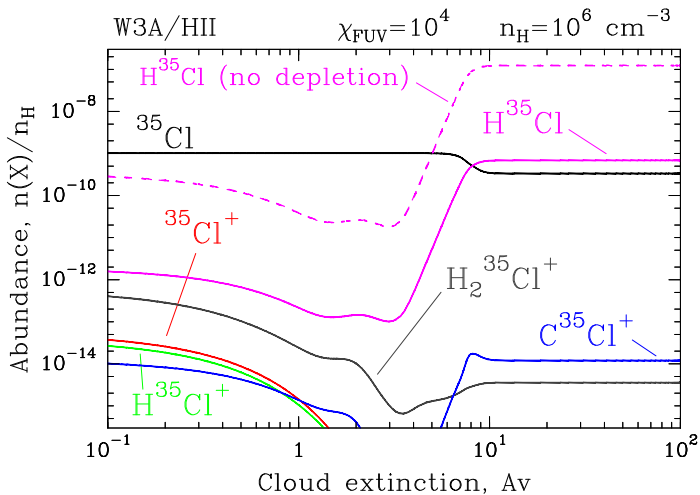}}
\caption{Cloud-depth dependent photochemical models adapted to W3\,A physical conditions.
The gas density (n$_H$=10$^6$\,cm$^{-3}$) and temperature (T$_k$=100\,K) are kept constant.
The UV radiation field is $\chi$=10$^4$ and the  ionization rate due to cosmic rays
is $\zeta$(H)=2.5$\times$10$^{-17}$\,s$^{-1}$.
The predicted abundance of several Cl-bearing species are shown. 
A chlorine gas-phase abundance of 1$\times$10$^{-9}$ is used.
The dashed line shows the expected HCl abundance with an undepleted abundance of  [Cl]=1.8$\times$10$^{-7}$.}
\label{fig:pdr-model}
\end{figure}

\subsection{Chlorine chemistry in W3\,A H{\sc ii} region}
\label{chemistry}

In the UV-illuminated gas, the chlorine chemistry involves
the reaction of Cl$^+$ with molecular hydrogen to form HCl$^+$,
which then reacts with H$_2$ to produce H$_2$Cl$^+$. If the
abundance of electrons is high, dissociative recombination of
H$_2$Cl$^+$ leads to the formation of HCl with a typically assumed
branching ratio of $\sim$10\%. 
In cloud interiors, atomic chlorine is mostly neutral, not ionized, so that the reaction 
of Cl and H$_3^+$ drives the formation of H$_2$Cl$^+$, which 
then reacts with CO and H$_2$O 
leading to  the formation of HCl. An alternative direct route to HCl in either
hot gas or  regions where vibrationally excited H$_2$ is abundant
(\textit{e.g.,} Ag\'undez et al. 2010) is the
Cl~+~H$_2$~$\rightarrow$~HCl~+~H reaction, which possesses an
energy barrier of $\sim$0.2\,eV \citep{Dobis2002}. The destruction
of HCl is dominated both by photoionization and photodissociation and
by reactions with C$^+$ and H$_{3}^{+}$ (the latter in the UV
shielded gas).
Previous observational studies of HCl suggest that there has been
a depletion of gas--phase chlorine
in dense molecular clouds of a factor of $>$100 (SFL96) relative to the elemental chlorine 
abundance observed in diffuse clouds ([Cl]$\simeq$1.8$\times$10$^{-7}$; 
\citet{Savage1996}, \citet{Sonnentrucker2006} and references therein).
To follow the HCl chemistry in the particular environment
of W3\,A H{\sc ii} region and  estimate
the Cl depletion we modeled the Cl-photochemistry
using the Meudon PDR code (Le Petit et al. 2006; Goicoechea \& Le
Bourlot 2007). The reaction network for Cl-bearing molecules
includes the updated rates of Neufeld \& Wolfire (2009). We note that
the X-ray luminosity reported in the region
($L_X$$\sim$5$\times$10$^{31}$~erg\,s$^{-1}$; Hofner et al. 2002)
is insufficient for a ``XDR--''  rather than a
``PDR--dominated''  environment. However, in cloud interiors
 X-ray photons may play an important role in the Cl-chemistry.

Figure \ref{fig:pdr-model} shows the output of a model adapted to
the physical conditions in W3\,A. The UV radiation field produced
by the OB stars in the region is simulated by an enhancement of
10$^4$ times the mean interstellar radiation field (in Draine
units). To reproduce the inferred H$^{35}$Cl abundance
($\sim$7.5$\times$10$^{-10}$), a gas-phase chlorine depletion of
$\lesssim$200 is needed, HCl accounting for $\approx$70\% of
the Cl nuclei in the gas phase. 
This conclusion is reached assuming
that the observed HCl arises in regions of large H$_2$ column
density (A$_V$$\gtrsim$100; see Sect.\ref{column_densities}), which is consistent with the submm
continuum maps of the region (Jaffe et al. 1983). If the observed
HCl arises in regions of lower extinction, the Cl depletion
factor will obviously be lower.

\section{Conclusions}
\label{sec:conclusions}

We have presented the first detection of H$^{35}$Cl and H$^{37}$Cl towards
the W3\,A H{\sc ii} region. 
The inferred H$^{35}$Cl column density (a few times
$\sim$10$^{14}$\,cm$^{-2}$) and fractional abundance
($\sim$7.5$\times$10$^{-10}$ per H nucleus) provide an
upper limit to the gas phase chlorine depletion of $\approx$200. This
value is lower than that observed towards Orion hot core, but
similar to that inferred towards Mon~R2 (SFL96). Radiative
transfer models including HFS line overlap and pumping by dust
photons have been used to interpret the observations. The best-fit
model provides a H$^{35}$Cl/H$^{37}$Cl abundance ratio of
$\approx$2.1, which is both lower than the solar
value ($\approx$3.1) and lower than the previous estimate towards
Orion ($\approx$4-6; SFL96). On the other hand, it is similar to
the [$^{35}$Cl]/[$^{37}$Cl] ratio obtained in the
IRC+10216 circumstellar envelope  from [Na$^{35}$Cl]/[Na$^{37}$Cl]
and [Al$^{35}$Cl]/[Al$^{37}$Cl] measurements \citep{Cernicharo1987,
Cernicharo2000}.

\begin{acknowledgements}
HIFI has been designed and built by a consortium of institutes and university departments from across
Europe, Canada and the United States under the leadership of SRON Netherlands Institute for Space
Research, Groningen, The Netherlands and with major contributions from Germany, France and the US.
Consortium members are: Canada: CSA, U.Waterloo; France: CESR, LAB, LERMA, IRAM; Germany:
KOSMA, MPIfR, MPS; Ireland, NUI Maynooth; Italy: ASI, IFSI-INAF, Osservatorio Astrofisico di Arcetri-
INAF; Netherlands: SRON, TUD; Poland: CAMK, CBK; Spain: Observatorio Astronómico Nacional (IGN),
Centro de Astrobiolog\'{\i}a (CSIC-INTA). Sweden: Chalmers University of Technology - MC2, RSS \& GARD;
Onsala Space Observatory; Swedish National Space Board, Stockholm University - Stockholm Observatory;
Switzerland: ETH Zurich, FHNW; USA: Caltech, JPL, NHSC.
We thank the Spanish MICINN for funding support through
grants AYA2006-14876, AYA2009-07304, and
and Consolider project CSD2009-00038.
JRG is supported by a \textit{Ram\'on y Cajal} research contract
from the Spanish MICINN. MA is supported by
a \textit{Marie Curie Intra-European Individual Fellowship} within
the EC FP7 under grant agreement n$^{\circ}$ 235753.
\end{acknowledgements}

\end{document}